\documentclass[sigconf]{acmart}

\AtBeginDocument{%
  }

\setcopyright{acmcopyright}
\copyrightyear{2024}
\acmYear{2024}
\acmDOI{XXXXXXX.XXXXXXX}

\acmConference[Conference acronym ICMAI]{Make sure to enter the correct
  conference title from your rights confirmation emai}{May 10--12,
  2024}{Beijing, China}
\acmPrice{15.00}
\acmISBN{978-1-4503-XXXX-X/18/06}

\begin{document}

\title{Predicting Mitral Valve mTEER Surgery Outcomes Using Machine Learning and Deep Learning Techniques}

\author{Tejas Vyas}
\affiliation{%
  \institution{Toronto Metropolitan University}
  \city{Toronto}
  \state{ON}
  \country{Canada}
}

\author{Mohsena Chowdhury}
\affiliation{%
  \institution{Toronto Metropolitan University}
  \city{Toronto}
  \state{ON}
  \country{Canada}
}

\author{Xiaojiao Xiao}
\affiliation{%
  \institution{Toronto Metropolitan University}
  \city{Toronto}
  \state{ON}
  \country{Canada}
}

\author{Mathias Claeys}
\affiliation{%
  \institution{St. Michael's Hospital - University of Toronto}
  \city{Toronto}
  \state{ON}
  \country{Canada}
}

\author{Géraldine Ong}
\affiliation{%
  \institution{St. Michael's Hospital - University of Toronto}
  \city{Toronto}
  \state{ON}
  \country{Canada}
}

\author{Guanghui Wang}
\affiliation{%
  \institution{Toronto Metropolitan University}
  \city{Toronto}
  \state{ON}
  \country{Canada}}
  \email{wangcs@torontomu.ca}

\renewcommand{\shortauthors}{Vyas et al.}

\begin{abstract}
Mitral Transcatheter Edge-to-Edge Repair (mTEER) is a medical procedure utilized for the treatment of mitral valve disorders. However, predicting the outcome of the procedure poses a significant challenge. This paper makes the first attempt to harness classical machine learning (ML) and deep learning (DL) techniques for predicting mitral valve mTEER surgery outcomes. To achieve this, we compiled a dataset from 467 patients, encompassing labeled echocardiogram videos and patient reports containing Transesophageal Echocardiography (TEE) measurements detailing Mitral Valve Repair (MVR) treatment outcomes. Leveraging this dataset, we conducted a benchmark evaluation of six ML algorithms and two DL models. The results underscore the potential of ML and DL in predicting mTEER surgery outcomes, providing insight for future investigation and advancements in this domain.
\end{abstract}


\maketitle

\section{Introduction}
\label{sec:intro}

The mitral valve, one of four valves in the heart that keep blood flowing in the right direction, plays a pivotal role in the intricate orchestration of blood flow between the left atrium and the left ventricle. Among the spectrum of cardiac disorders, mitral valve prolapse stands out as a notable concern, often progressing to the widespread condition known as Mitral Valve Regurgitation (MVR). Particularly prevalent among the elderly population, MVR presents a significant health challenge, severely impacting physiological functions and often necessitating surgical intervention \cite{badhwar2023risk}.

Mitral Transcatheter Edge-to-Edge Repair (mTEER) emerges as a beacon of hope in the realm of cardiac surgery, offering a less invasive alternative to traditional open-heart procedures. This innovative surgical approach has proven to be a viable and effective option for patients dealing with MVR \cite{camaj2023heart}. By avoiding the complexities and risks associated with open-heart surgery, mTEER minimizes the invasiveness of the intervention, promoting faster recovery and reduced postoperative complications.
However, the success of the mTEER procedure is intricately tied to the unique anatomy of the mitral valve. As such, accurately predicting the outcomes of this surgery becomes paramount, not only for optimizing the allocation of medical resources but also for tailoring the approach to ensure personalized and effective patient care. The ability to anticipate the nuances of each individual mitral valve's anatomy enhances the precision and efficacy of the mTEER procedure, contributing to improved overall patient outcomes and quality of life.



Traditional analysis of patient reports and echocardiogram (ECG) videos by cardiologists is resource-intensive. In contrast, our approach capitalizes on traditional machine learning (ML) and deep learning (DL) techniques for outcome prediction. Notably, only a limited number of artificial intelligence studies have addressed this prediction challenge, and even fewer have incorporated ECG video data \cite{gheorghe}\cite{ penso}. This paper makes strides towards enhancing patient care through technological innovations and demonstrates the possibility of predicting patient outcomes based on ECG videos and patient reports based on DL and ML technologies.

The major contributions of this study include (i) We curated a dataset from 467 patients who underwent mitral valve mTEER surgery, labeling the data with both echocardiogram videos and patient reports containing Transesophageal Echocardiography (TEE) measurements. This dataset marks the first of its kind in this field; (ii) Through an in-depth exploration of the dataset, we conducted a benchmark study involving six ML algorithms and two DL models to predict the outcomes of mTEER surgery; and (iii) The insights gleaned from this study establish a baseline for mTEER surgery prediction, serving as a foundation for further refinement and optimization of performance through subsequent research.

\section{Related Work}
\label{sec:related}

Predictive modeling for surgical outcomes has become a focal point of attention in recent years, particularly in clinical settings such as Cardiology \cite{jalali2020deep}. Understanding the necessity of outcome prediction is pivotal for achieving optimal patient management. A noteworthy contribution to this field comes from the work of Geyer et al. \cite{geyer}, who conducted a comprehensive analysis of the factors influencing survival in Edge-to-Edge Repair techniques. Their findings highlight the considerable potential of predictive models in enhancing the support provided for patient care.

Machine learning and deep learning techniques have been widely applied in the medical field for image classification \cite{patel2022discriminative}\cite{yang2022unsupervised}\cite{zhang2023gender}, lesion detection \cite{bur2023interpretable}\cite{li2021colonoscopy}\cite{wilson2022harnessing}, tumor segmentation \cite{huo2018supervoxel}\cite{patel2022fuzzynet}\cite{xiao2023edge}, etc. However, very few studies have been reported in the context of Mitral Valve Repair methods, Penso et al. \cite{penso} conducted a study leveraging Machine Learning to predict MVR and the recurrence of Mitral Regurgitation in patients post-surgery. Despite the innovative approach, their focus was on general surgical repair methods and did not target percutaneous techniques such as mTEER. Additionally, their research did not encapsulate the potential contributions of imaging data in patient outcome prediction.

The significance of imaging, especially echocardiography, in diagnosing and managing MVR has been previously documented in studies such as the article by Gheorghe et al. \cite{gheorghe}. Their study illustrated the importance of echocardiography as an interventional imaging technique and underscores the potential of understanding the physical abnormalities of the mitral valve through the imaging technique, subsequently aiding in surgical planning and assessing intervention effectiveness.

The transformative impact of technological advancements on the management and diagnosis of MVR is evident as novel imaging and analysis techniques continue to emerge. Recognizing the significance of these developments, our study aims to bridge the gap between traditional methods and the opportunities presented by state-of-the-art imaging and machine learning techniques. The focal point of our investigation is directed towards assessing the outcomes of mTEER surgery within the context of these advancements.

\section{Methodology}
\label{sec:method}

\subsection{Dataset Generation}

We collected the data from 467 patients at St. Michael's Hospital in Toronto. With the help of clinical experts, the data of each patient is composed of both echocardiogram videos and patient reports containing TEE measurements that elaborate on MVR treatment outcomes. The initial patient data contains 53 features, covering demographics, procedural details, valve dimensions, and regurgitation characterizations. After excluding non-contributory features and those with null values, we refined the dataset to 43 predictive features. Numerical features remained unchanged, while categorical features underwent encoding through one-hot encoding.

\begin{figure}[t]
    \centering
    \includegraphics[width=1\linewidth]{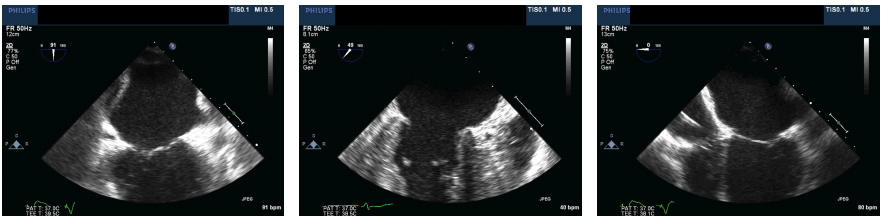}
    \caption{Sample ECG frames of 2-valve (left), 3-valve (middle), and 4-valve (right).}
    \label{fig:1}
\end{figure}

The video dataset comprised 1,374 echocardiogram clips, segmented by valve count for the original 467 patients. In preparation for training the deep learning model, we converted the video clips into image frames, generating a total of 72,952 images. Each patient typically contributed three clips corresponding to 2-valve, 3-valve, and 4-valve echocardiograms, as illustrated in Fig. \ref{fig:1}. Owing to varying image equipment settings, the video clips have 13 distinct resolutions, with heights ranging from 708 to 894 pixels and widths ranging from 948 to 1588 pixels. 

The data of the original 467 patients were labeled to indicate procedural success or failure, yielding approximately 65\% success labels and 35\% failure labels. This paper employs a 10-fold cross-validation approach. The dataset is randomly divided into 10 equal groups, with particular attention paid to preventing data leakage. The data is split at the patient level, ensuring that frames from the same patient do not appear in both the training and testing sets. This meticulous split methodology is designed to prevent overestimation of the model's performance and guarantees that the trained model generalizes effectively to new patients. Additionally, the testing fold used was kept the same in both ML and DL experiments to allow comparative analysis of both techniques.

\subsection{Model Selection}
The study aims to provide a benchmark evaluation of mTEER prediction based on both ML algorithms and deep DL models. Specifically, we evaluated the following six ML algorithms: Decision Tree (DT), Logistic Regression (LR), k-Nearest Neighbors (kNN), Random Forest (RF), Gradient Boost (GB), Support Vector Machine (SVM). Additionally, we chose two baseline DL models: VGG16 with modified final layers and EfficientNet-B0 \cite{tan2019efficientnet}.

\subsection{Pre-processing steps}

\textbf{Machine Learning.}
For ML modelling, we leverage the Scikit-learn library\cite{scikit-learn} to construct the classification model following several pre-processing steps, which include:
\begin{itemize}
    \item Importing necessary libraries and spreadsheet data.
    \item Conducting data pre-processing, encompassing feature selection and data cleaning, to obtain the final data frame.
    \item Generating cross-validation folds to acquire training and testing sets.
    \item Implementing variables to store hyperparameter ranges for each model, facilitating their utilization during Grid Search.
\end{itemize}

\textbf{Deep Learning.}
For DL modelling, we leverage several DL libraries including the PyTorch library\cite{pytorch} to construct the deep learning models following several preprocessing steps. To mitigate overfitting, the following measures were implemented:
\begin{itemize}
    \item Initialization of pre-trained models with variants supporting Batch Normalization.

    \item Application of data augmentation techniques, encompassing resizing, random flips, rotations, and normalization.
    \item Incorporation of early stopping support to facilitate optimal convergence.
\end{itemize}

\subsection{Model Training}
\textbf{Machine Learning.}
For Machine Learning model training, parameter ranges were provided to the grid search function with all random seed values set to 21. The best parameters determined through the grid search were subsequently employed to train the candidate models, and these optimal hyperparameters are detailed in Table~\ref{tab:1}.

\begin{table}
  \caption{Hyperparameters of Learning Models}
  \label{tab:1}
  \begin{tabular}{ll}
    \toprule
    \textbf{Model}&\textbf{Parameter Value}\\
    \midrule
    \textbf{Decision Tree} & Criterion = 'gini' \\
 & Max Depth = None \\
 & Min Samples Leaf = 10 \\
 & Min Samples Split = 2 \\
  \hline
  \textbf{Logistic Regression} & C = 0.001 \\
 & Penalty = 'none' \\
 & Solver = 'sag' \\
  \hline
\textbf{Gradient Boosting} & Learning Rate = 0.01 \\
 & Max Depth = 5 \\
 & Max Features = 'log2' \\
 & N Estimators = 200 \\
 & Subsample = 0.7 \\
  \hline
\textbf{Random Forest} & Bootstrap = True \\
 & Max Depth = 5 \\
 & Min Samples Leaf = 1 \\
 & Min Samples Split = 8 \\
 & N Estimators = 100 \\
  \hline
\textbf{K-Nearest Neighbors} & Metric = 'euclidean' \\
 & N Neighbors = 30 \\
 & Weights = 'uniform' \\
  \hline
\textbf{Support Vector Machine} & C = 1 \\
 & Gamma = 0.001 \\
 & Kernel = 'poly' \\
 \hline
 \hline
 \textbf{VGG16} & 
Batch Size: 128 \\
& Loss: NLLLoss \\
& Max Epochs: 10 \\
& Starting Learning Rate: 0.001 \\
& Maximum Learning Rate: 0.15 \\
& Optimizer: SGD \\
\hline
\textbf{EfficientNet-B0} & 
Batch Size: 128 \\
& Loss: NLLLoss \\
& Max Epochs: 10 \\
& Starting Learning Rate: 0.001 \\
& Maximum Learning Rate: 0.15 \\
& Optimizer: RMSProp \\
  \bottomrule
\end{tabular}
\end{table}

\textbf{Deep Learning.}
In the generation of deep learning prototypes, we established the following experimental setup, conducting training with 10-fold cross-validation utilizing Nvidia 3090 GPUs. Table \ref{tab:1} outlines the hyperparameters employed in the experimentation for all deep learning techniques.

Given that VGG16 was initially designed for segmentation, the final layer of the model was modified to accommodate binary classification. The adjusted architecture is as follows:
\begin{itemize}
    \item Input Layer: 4096 units
    \item Hidden Layer 1: Linear, 4096 to 256 units
    \item Activation Function: ReLU
    \item Regularization Layer: Dropout, rate = 0.4
    \item Hidden Layer 2: Linear, 256 to 2 units
    \item Output Layer: LogSoftmax, dimension = 1
\end{itemize}

\subsection{Performance Metrics}
The following metrics were adopted for both machine learning and deep learning models \cite{li2021colonoscopy}\cite{mcclannahan2021classification}:

\textbf{Confusion Matrix:} Confusion matrix is a table used to evaluate the performance of a classification model.
\[
\begin{array}{c|cc}
 & \text{Predicted: Yes} & \text{Predicted: No} \\
\hline
\text{Actual: Yes} & TP & FN \\
\text{Actual: No} & FP & TN \\
\end{array}
\]
Here TP and TN stand for True Positive and True Negative, and FP and FN stand for False Positive and False Negative, respectively. Confusion matrix is particularly useful for understanding the types and frequencies of errors that a model makes.

\textbf{Accuracy (Frame Accuracy):} Measures the overall correctness of the model, i.e., the ratio of correctly predicted instances to the total number of instances.
\begin{equation}
\text{Accuracy} = \frac{TP + TN}{TP + TN + FP + FN}
\label{eq:accuracy}
\end{equation}

\textbf{Patient Level Accuracy}: A new performance metric which uses frame level accuracy and calculates the majority result per patient to generate patient-level accuracy. Specifically, we calculate the values of TP, TN, FP, FN in Eq.\eqref{eq:accuracy} by aggregating all respective frame predictions \(p\) as below.
\begin{equation}
M_i = 
\begin{cases} 
1 & \text{if} \;\;\frac{\sum_{i=1}^{n} p_i}{n} > 0.5 \\
0 & \text{otherwise}
\end{cases}
\end{equation}
where \(n\) is the number of patients, and \(M_i\) is the majority prediction.

\textbf{F1-Score:} F1 score is a metric that combines precision and recall into a single value, providing a balance between these two performance measures. It is particularly useful in situations where there is an imbalance between the number of positive and negative instances in the dataset.
\begin{equation}
\text{F1-Score} = \frac{2 \times \text{Precision} \times \text{Recall}}{\text{Precision} + \text{Recall}}
\label{eq:f1score}
\end{equation}
where ${Precision} = \frac{TP}{TP + FP}$ and ${Recall} = \frac{TP}{TP + FN}$.

\textbf{AUC:} Area under the receiver operating characteristic (ROC) curve, which is a graphical representation that illustrates the trade-off between sensitivity (True Positive Rate) and specificity (True Negative Rate) across different thresholds.
\begin{equation}
\text{AUC} = \sum_{i} \text{TPR}(t_i) \times \Delta \text{FPR}(t_i)
\label{eq:auc}
\end{equation}
where $TPR =  = \frac{TP}{TP + FN}$ and $FPR = \frac{FP}{FP + TN}$.

\section{Results and Discussion}
\label{sec:results}

\subsection{Machine Learning}
The results obtained for different machine learning algorithms are shown in Tables \ref{tab:acc} and \ref{tab:con}. To save space, we organize all confusion matrices in a tabular format.
It is evident from the results that Logistic Regression (LR) demonstrated the most favorable results among all traditional ML models, closely followed by Support Vector Machine. Random Forest, Gradient Boosting, and k-nearest neighbors exhibited performance that was nearly on par with SVM in terms of training accuracy and AUC score. The strong performance of LR suggests the presence of a potential linear relationship within the features, effectively capturing the underlying patterns associated with surgical outcomes in the specific case. However, the marginal differences in performance among LR and the other models underscore the complexity of the problem domain. In such cases, hyperparameter tuning becomes pivotal, in addition to the influence of the randomly generated folds during cross-validation.

\begin{table}
  \caption{Accuracy and evaluation metrics for ML models.}
  \label{tab:acc}
  \begin{tabular}{lcccc}
    \toprule
    Model & Training Acc. & Testing Acc. & AUC & F1 \\
    \midrule
DT & 0.591$\pm 0.07$ & 0.534$\pm 0.08$ & 0.474 & 0.650 \\
\textbf{LR} & \textbf{0.673$\pm 0.05$} & \textbf{0.661$\pm 0.04$} & \textbf{0.555} & \textbf{0.777} \\
GB & 0.646$\pm 0.06$ & 0.620$\pm 0.06$ & 0.498 & 0.756 \\
RF & 0.649$\pm 0.06$ & 0.624$\pm 0.07$ & 0.532 & 0.747 \\
kNN & 0.640$\pm 0.05$ & 0.612$\pm 0.05$ & 0.505 & 0.744 \\
SVM & 0.661$\pm 0.04$ & 0.642$\pm 0.05$ & 0.553 & 0.757 \\
  \bottomrule
\end{tabular}
\end{table}

\begin{table}
  \caption{Confusion matrix for ML models.}
  \label{tab:con}
  \begin{tabular}{lcccc}
    \toprule
    Model & FP & TP & FN & TN \\
    \midrule
DT & 12 & 20 & 10 & 5 \\
\textbf{LR} & \textbf{13} & \textbf{28} & \textbf{2} & \textbf{3} \\
GB & 15 & 28 & 3 & 1 \\
RF & 13 & 26 & 4 & 3 \\
kNN & 14 & 26 & 4 & 2 \\
SVM & 13 & 26 & 4 & 4 \\
  \bottomrule
\end{tabular}
\end{table}

\subsection{Deep Learning}
The accuracy for the deep learning models at both frame level and patient level is shown in Table \ref{tab:dlacc}. The AUC and F1 scores are shown in Table \ref{tab:auc}. The confusion matrix at both frame level and patient level is shown in Table \ref{tab:dlcon}. The results are obtained via 10-fold cross-validation, i.e., the model is trained and evaluated 10 times, each time using a different fold as the test set and the remaining nine folds as the training set. This process is repeated until each fold has been used as the test set exactly once. As we see from the results, VGG16 performs better than EfficientNet-B0 on all metrics, with the patient-wise accuracy is about 3\% higher than frame-wise accuracy, with the exception of the AUC score, where both models demonstrate nearly equivalent performance. 

\begin{table}
  \caption{Training and Accuracy for DL models.}
  \label{tab:dlacc}
  \begin{tabular}{lccc}
    \toprule
    Model & Train Acc. & Frame Acc. & Patient Acc. \\
    \midrule
\textbf{VGG16} & \textbf{0.671$\pm 0.04$} & \textbf{0.630$\pm 0.05$} & \textbf{0.664$\pm 0.04$} \\
EfficientNet-B0 & 0.578$\pm 0.04$ & 0.615$\pm 0.07$ & 0.637$\pm 0.07$ \\
  \bottomrule
\end{tabular}
\end{table}
\begin{table}
  \caption{AUC and F1 Score for DL models.}
  \label{tab:auc}
  \begin{tabular}{lcc}
    \toprule
    Model & AUC & F1 \\
    \midrule
\textbf{VGG16} & 0.524 & \textbf{0.757} \\
EfficientNet-B0 & \textbf{0.526} & 0.725 \\
  \bottomrule
\end{tabular}
\end{table}
\begin{table}
  \caption{Confusion matrix for DL models (frame level/patient level).}
  \label{tab:dlcon}
  \begin{tabular}{lcccc}
    \toprule
    Model & FP & TP & FN & TN \\
    \midrule
\textbf{VGG16} & \textbf{2304/12} & \textbf{4319/28} & \textbf{386/2} & \textbf{286/3} \\
EfficientNet-B0 & 2246/15 & 4154/22 & 551/2 & 345/6 \\
  \bottomrule
\end{tabular}
\end{table}




\subsection{Comparative Analysis}
Throughout the study, two distinct datasets were utilized. The first dataset consisted of a spreadsheet containing patient metrics designed for traditional ML applications. The second dataset comprised images extracted from echocardiogram videos, specifically tailored for DL applications. While both datasets originated from the same cohort of patients, the nature and methodologies applied to each dataset exhibited significant variations.

The ML dataset, consisting of patient data, underwent training using the grid search method, incorporating a 10-fold cross-validation after partitioning a 20\% test set to enhance result robustness. After discovering the best set of hyperparameters from the grid search, the final results for the respective ML models were extracted using a straightforward 10-fold cross-validation with the tenth fold, representing 10\% of the data, reserved for testing purposes. The DL dataset comprising images underwent a similar stratified 10-fold cross-validation, and the patient cohort within the tenth fold was kept consistent across both ML and DL experiments to facilitate comparative analysis.

Owing to the variations in methodology and the distinct natures of the datasets, direct comparisons between the metrics of the DL and ML models were constrained. Nonetheless, based on current results, both models provided comparable outcomes, with DL providing superior accuracy, and ML providing better AUC and F1 scores. 
The DL results, derived from image data, offer insights specific to visual patterns in the ECG video recordings. In contrast, the ML results, grounded in quantifiable patient metrics, provide an understanding of clinical patterns based on the measurements within the dataset.

ML techniques excel at processing structured data, such as the tabulated structure of patient measurements. Their explainability and efficiency with smaller datasets allow for the extraction of meaningful patterns, even with relatively modest datasets. However, it is time-consuming and demands domain expertise to measure and interpret the data. In contrast, DL techniques showcase greater prowess in discerning patterns within complex data, particularly images. The robust nature of Convolutional Neural Networks (CNNs) contributes to effective overfitting prevention and better handling of larger datasets. The choice between ML and DL techniques may therefore hinge on factors such as data set availability and system constraints, including computational resources, rather than solely relying on a direct comparison of accuracy or other metrics.

Given sufficient data, DL techniques generally outperform ML approaches. However, in this study, both techniques demonstrate similar performance, primarily attributed to the limited size of the dataset. To further enhance the performance, one approach could be leveraging ensemble methods capable of handling multi-modal data. Additionally, conducting further experimentation by incorporating both report data and video frames for prediction could be explored. 

\subsection{Discussion}
Video classification poses a non-trivial challenge, and a significant obstacle encountered was the pre-processing of video data. Given the absence of a widely used dataset specifically dedicated to mTEER ECG images, a meticulous process was undertaken to convert the videos into frames while maintaining a labeling scheme that included vital information.
While this study currently focuses on a binary classification problem due to dataset labeling, it's essential to note that in the clinical field, assessments of surgery outcomes are not binary but are typically rated categorically. 

For example, sometimes it is hard to classify the surgery outcomes as either success or failure. It is reasonable to add more categories in between. This will involve more domain expertise to create a more comprehensive fine-tuned dataset that encompasses all categories, allowing for multi-class prediction. This can be particularly valuable in the clinical space when collaborating with a cardiology domain expert. Moreover, mTEER images include other categories like Doppler coloring and 3D views. Future experiments could explore modeling with these images, providing an opportunity for a more nuanced understanding and prediction. This updated dataset has the potential to offer more context to the deep learning model, enhancing its capabilities.



\section{Conclusion}

In this study, we have curated two datasets for predicting mTEER surgery outcomes. Through meticulous data preprocessing and the application of both ML and DL techniques, comprehensive insights have been gained into surgery outcomes. This enables the study to offer robust baseline results, including a thorough comparative analysis of various methods. This study is a pioneer in utilizing mTEER ECG videos for classification and outcome prediction through deep learning methods, achieving commendable accuracy and robustness.
The findings from this study establish a robust foundation for future research in an area with limited current exploration. This groundwork enables more effective predictions for surgical outcomes and establishes a baseline that can be further refined and optimized through ongoing experimentation and research.

An analysis of the potential of employing these methods for cardiac prediction, along with an exploration of the numerous challenges involved, opens up various opportunities for future research in harnessing multimodal data for surgical outcome predictions, particularly for mTEER patients. The insights gained from this study, coupled with ongoing experimentation and future research endeavors, have the potential to substantially improve predictive accuracy. Consequently, these advancements can make valuable contributions to the field of medical diagnostics and treatment planning.

\section*{Acknowledgement}
This work was partly supported by the Natural Sciences and
Engineering Research Council of Canada (NSERC) and TMU FOS Dean's Research Fund.

\balance
{
\bibliographystyle{ACM-Reference-Format}
\bibliography{egbib}


\begin{thebibliography}{19}


\ifx \showCODEN    \undefined \def \showCODEN     #1{\unskip}     \fi
\ifx \showDOI      \undefined \def \showDOI       #1{#1}\fi
\ifx \showISBNx    \undefined \def \showISBNx     #1{\unskip}     \fi
\ifx \showISBNxiii \undefined \def \showISBNxiii  #1{\unskip}     \fi
\ifx \showISSN     \undefined \def \showISSN      #1{\unskip}     \fi
\ifx \showLCCN     \undefined \def \showLCCN      #1{\unskip}     \fi
\ifx \shownote     \undefined \def \shownote      #1{#1}          \fi
\ifx \showarticletitle \undefined \def \showarticletitle #1{#1}   \fi
\ifx \showURL      \undefined \def \showURL       {\relax}        \fi
\providecommand\bibfield[2]{#2}
\providecommand\bibinfo[2]{#2}
\providecommand\natexlab[1]{#1}
\providecommand\showeprint[2][]{arXiv:#2}

\bibitem[Badhwar et~al\mbox{.}(2023)]%
        {badhwar2023risk}
\bibfield{author}{\bibinfo{person}{Vinay Badhwar}, \bibinfo{person}{Joanna Chikwe}, \bibinfo{person}{A~Marc Gillinov}, {and} \bibinfo{person}{et~al. Vemulapalli}.} \bibinfo{year}{2023}\natexlab{}.
\newblock \showarticletitle{Risk of surgical mitral valve repair for primary mitral regurgitation}.
\newblock \bibinfo{journal}{\emph{Journal of the American College of Cardiology}} \bibinfo{volume}{81}, \bibinfo{number}{7} (\bibinfo{year}{2023}), \bibinfo{pages}{636--648}.
\newblock


\bibitem[Bur et~al\mbox{.}(2023)]%
        {bur2023interpretable}
\bibfield{author}{\bibinfo{person}{Andr{\'e}s~M Bur}, \bibinfo{person}{Tianxiao Zhang}, \bibinfo{person}{Xiangyu Chen}, \bibinfo{person}{Hannah Kavookjian}, {and} \bibinfo{person}{et al.}} \bibinfo{year}{2023}\natexlab{}.
\newblock \showarticletitle{Interpretable Computer Vision to Detect and Classify Structural Laryngeal Lesions in Digital Flexible Laryngoscopic Images}.
\newblock \bibinfo{journal}{\emph{Otolaryngology--Head and Neck Surgery}} (\bibinfo{year}{2023}).
\newblock


\bibitem[Camaj et~al\mbox{.}({[n.\,d.]})]%
        {camaj2023heart}
\bibfield{author}{\bibinfo{person}{Anton Camaj}, \bibinfo{person}{Vinod~H Thourani}, \bibinfo{person}{Linda~D Gillam}, {and} \bibinfo{person}{Gregg~W Stone}.} \bibinfo{year}{[n.\,d.]}\natexlab{}.
\newblock \showarticletitle{Heart Failure and Secondary Mitral Regurgitation: A Contemporary Review}.
\newblock \bibinfo{journal}{\emph{Journal of the Society for Cardiovascular Angiography \& Interventions}} \bibinfo{volume}{2}, \bibinfo{number}{6} (\bibinfo{year}{[n.\,d.]}).
\newblock


\bibitem[Geyer et~al\mbox{.}(2021)]%
        {geyer}
\bibfield{author}{\bibinfo{person}{M. Geyer}, \bibinfo{person}{K. Keller}, \bibinfo{person}{S. Born}, \bibinfo{person}{K. Bachmann}, \bibinfo{person}{A.~R. Tamm}, \bibinfo{person}{T.~F. Ruf}, \bibinfo{person}{F. Kreidel}, \bibinfo{person}{O. Hahad}, \bibinfo{person}{M. Ahoopai}, \bibinfo{person}{L. Hobohm}, \bibinfo{person}{A. Beiras-Fernandez}, \bibinfo{person}{A. Kornberger}, \bibinfo{person}{E. Schulz}, \bibinfo{person}{T. Münzel}, {and} \bibinfo{person}{R.~S. von Bardeleben}.} \bibinfo{year}{2021}\natexlab{}.
\newblock \showarticletitle{Predictors of short- and long-term outcomes of patients undergoing transcatheter mitral valve edge-to-edge repair}.
\newblock \bibinfo{journal}{\emph{Catheterization and Cardiovascular Interventions}} \bibinfo{volume}{97}, \bibinfo{number}{3} (\bibinfo{date}{2} \bibinfo{year}{2021}), \bibinfo{pages}{E390--E401}.
\newblock


\bibitem[Gheorghe et~al\mbox{.}(2021)]%
        {gheorghe}
\bibfield{author}{\bibinfo{person}{L.~L. Gheorghe}, \bibinfo{person}{S. Mobasseri}, \bibinfo{person}{E. Agricola}, \bibinfo{person}{D.~D. Wang}, \bibinfo{person}{F. Milla}, \bibinfo{person}{M. Swaans}, {and} \bibinfo{person}{et al.}} \bibinfo{year}{2021}\natexlab{}.
\newblock \showarticletitle{Imaging for Native Mitral Valve Surgical and Transcatheter Interventions}.
\newblock \bibinfo{journal}{\emph{JACC: Cardiovascular Imaging}} \bibinfo{volume}{14}, \bibinfo{number}{1} (\bibinfo{date}{1} \bibinfo{year}{2021}), \bibinfo{pages}{112--127}.
\newblock


\bibitem[Huo et~al\mbox{.}(2018)]%
        {huo2018supervoxel}
\bibfield{author}{\bibinfo{person}{Jie Huo}, \bibinfo{person}{Jonathan Wu}, \bibinfo{person}{Jiuwen Cao}, {and} \bibinfo{person}{Guanghui Wang}.} \bibinfo{year}{2018}\natexlab{}.
\newblock \showarticletitle{Supervoxel based method for multi-atlas segmentation of brain MR images}.
\newblock \bibinfo{journal}{\emph{NeuroImage}}  \bibinfo{volume}{175} (\bibinfo{year}{2018}), \bibinfo{pages}{201--214}.
\newblock


\bibitem[Jalali et~al\mbox{.}(2020)]%
        {jalali2020deep}
\bibfield{author}{\bibinfo{person}{Ali Jalali}, \bibinfo{person}{Hannah Lonsdale}, \bibinfo{person}{Nhue Do}, \bibinfo{person}{Jacquelin Peck}, \bibinfo{person}{Monesha Gupta}, {and} \bibinfo{person}{et al.}} \bibinfo{year}{2020}\natexlab{}.
\newblock \showarticletitle{Deep learning for improved risk prediction in surgical outcomes}.
\newblock \bibinfo{journal}{\emph{Scientific reports}} \bibinfo{volume}{10}, \bibinfo{number}{1} (\bibinfo{year}{2020}), \bibinfo{pages}{9289}.
\newblock


\bibitem[Li et~al\mbox{.}(2021)]%
        {li2021colonoscopy}
\bibfield{author}{\bibinfo{person}{Kaidong Li}, \bibinfo{person}{Mohammad~I Fathan}, \bibinfo{person}{Krushi Patel}, {and} \bibinfo{person}{et al}.} \bibinfo{year}{2021}\natexlab{}.
\newblock \showarticletitle{Colonoscopy polyp detection and classification: Dataset creation and comparative evaluations}.
\newblock \bibinfo{journal}{\emph{Plos one}} \bibinfo{volume}{16}, \bibinfo{number}{8} (\bibinfo{year}{2021}), \bibinfo{pages}{e0255809}.
\newblock


\bibitem[McClannahan et~al\mbox{.}(2021)]%
        {mcclannahan2021classification}
\bibfield{author}{\bibinfo{person}{Brian McClannahan}, \bibinfo{person}{Cucong Zhong}, {and} \bibinfo{person}{Guanghui Wang}.} \bibinfo{year}{2021}\natexlab{}.
\newblock \showarticletitle{Classification of Long Noncoding RNA Elements Using Deep Convolutional Neural Networks and Siamese Networks}.
\newblock \bibinfo{journal}{\emph{arXiv preprint arXiv:2102.05582}} (\bibinfo{year}{2021}).
\newblock


\bibitem[Paszke et~al\mbox{.}(2019)]%
        {pytorch}
\bibfield{author}{\bibinfo{person}{Adam Paszke}, \bibinfo{person}{Sam Gross}, \bibinfo{person}{Francisco Massa}, \bibinfo{person}{Adam Lerer}, \bibinfo{person}{James Bradbury}, {and} \bibinfo{person}{et al.}} \bibinfo{year}{2019}\natexlab{}.
\newblock \showarticletitle{PyTorch: An Imperative Style, High-Performance Deep Learning Library}. In \bibinfo{booktitle}{\emph{Advances in Neural Information Processing Systems 32}}, \bibfield{editor}{\bibinfo{person}{H.~Wallach}, \bibinfo{person}{H.~Larochelle}, \bibinfo{person}{A.~Beygelzimer}, \bibinfo{person}{F.~d'Alché Buc}, \bibinfo{person}{E.~Fox}, {and} \bibinfo{person}{R.~Garnett}} (Eds.). \bibinfo{publisher}{Curran Associates, Inc.}, \bibinfo{pages}{8024--8035}.
\newblock


\bibitem[Patel and Wang(2022)]%
        {patel2022discriminative}
\bibfield{author}{\bibinfo{person}{Krushi Patel} {and} \bibinfo{person}{Guanghui Wang}.} \bibinfo{year}{2022}\natexlab{}.
\newblock \showarticletitle{A discriminative channel diversification network for image classification}.
\newblock \bibinfo{journal}{\emph{Pattern Recognition Letters}}  \bibinfo{volume}{153} (\bibinfo{year}{2022}), \bibinfo{pages}{176--182}.
\newblock


\bibitem[Patel et~al\mbox{.}(2022)]%
        {patel2022fuzzynet}
\bibfield{author}{\bibinfo{person}{Krushi~Bharatbhai Patel}, \bibinfo{person}{Fengjun Li}, {and} \bibinfo{person}{Guanghui Wang}.} \bibinfo{year}{2022}\natexlab{}.
\newblock \showarticletitle{Fuzzynet: A fuzzy attention module for polyp segmentation}. In \bibinfo{booktitle}{\emph{NeurIPS'22 Workshop on All Things Attention: Bridging Different Perspectives on Attention}}.
\newblock


\bibitem[Pedregosa et~al\mbox{.}(2011)]%
        {scikit-learn}
\bibfield{author}{\bibinfo{person}{F. Pedregosa}, \bibinfo{person}{G. Varoquaux}, \bibinfo{person}{A. Gramfort}, \bibinfo{person}{V. Michel}, \bibinfo{person}{B. Thirion}, {and} \bibinfo{person}{et al.}} \bibinfo{year}{2011}\natexlab{}.
\newblock \showarticletitle{Scikit-learn: Machine Learning in {P}ython}.
\newblock \bibinfo{journal}{\emph{Journal of Machine Learning Research}}  \bibinfo{volume}{12} (\bibinfo{year}{2011}), \bibinfo{pages}{2825--2830}.
\newblock


\bibitem[Penso et~al\mbox{.}(2021)]%
        {penso}
\bibfield{author}{\bibinfo{person}{M. Penso}, \bibinfo{person}{M. Pepi}, \bibinfo{person}{V. Mantegazza}, \bibinfo{person}{C. Cefalù}, {and} \bibinfo{person}{et al.}} \bibinfo{year}{2021}\natexlab{}.
\newblock \showarticletitle{Machine Learning Prediction Models for Mitral Valve Repairability and Mitral Regurgitation Recurrence in Patients Undergoing Surgical Mitral Valve Repair}.
\newblock \bibinfo{journal}{\emph{Bioengineering}} \bibinfo{volume}{8}, \bibinfo{number}{9} (\bibinfo{date}{8} \bibinfo{year}{2021}), \bibinfo{pages}{117}.
\newblock


\bibitem[Tan and Le(2019)]%
        {tan2019efficientnet}
\bibfield{author}{\bibinfo{person}{Mingxing Tan} {and} \bibinfo{person}{Quoc Le}.} \bibinfo{year}{2019}\natexlab{}.
\newblock \showarticletitle{Efficientnet: Rethinking model scaling for convolutional neural networks}. In \bibinfo{booktitle}{\emph{International conference on machine learning}}. PMLR, \bibinfo{pages}{6105--6114}.
\newblock


\bibitem[Wilson et~al\mbox{.}(2022)]%
        {wilson2022harnessing}
\bibfield{author}{\bibinfo{person}{Blake~S Wilson}, \bibinfo{person}{Debara~L Tucci}, \bibinfo{person}{David~A Moses}, {and} \bibinfo{person}{et al.}} \bibinfo{year}{2022}\natexlab{}.
\newblock \showarticletitle{Harnessing the power of artificial intelligence in otolaryngology and the communication sciences}.
\newblock \bibinfo{journal}{\emph{Journal of the Association for Research in Otolaryngology}} \bibinfo{volume}{23}, \bibinfo{number}{3} (\bibinfo{year}{2022}), \bibinfo{pages}{319--349}.
\newblock


\bibitem[Xiao et~al\mbox{.}(2023)]%
        {xiao2023edge}
\bibfield{author}{\bibinfo{person}{Xiaojiao Xiao}, \bibinfo{person}{Qinmin~Vivian Hu}, {and} \bibinfo{person}{Guanghui Wang}.} \bibinfo{year}{2023}\natexlab{}.
\newblock \showarticletitle{Edge-aware multi-task network for integrating quantification segmentation and uncertainty prediction of liver tumor on multi-modality non-contrast MRI}. In \bibinfo{booktitle}{\emph{International Conference on Medical Image Computing and Computer-Assisted Intervention}}. \bibinfo{pages}{652--661}.
\newblock


\bibitem[Yang et~al\mbox{.}(2022)]%
        {yang2022unsupervised}
\bibfield{author}{\bibinfo{person}{Yiju Yang}, \bibinfo{person}{Tianxiao Zhang}, {and} \bibinfo{person}{et al.}} \bibinfo{year}{2022}\natexlab{}.
\newblock \showarticletitle{An unsupervised domain adaptation model based on dual-module adversarial training}.
\newblock \bibinfo{journal}{\emph{Neurocomputing}}  \bibinfo{volume}{475} (\bibinfo{year}{2022}), \bibinfo{pages}{102--111}.
\newblock


\bibitem[Zhang et~al\mbox{.}(2023)]%
        {zhang2023gender}
\bibfield{author}{\bibinfo{person}{Tianxiao Zhang}, \bibinfo{person}{Andr{\'e}s~M Bur}, \bibinfo{person}{Shannon Kraft}, {and} \bibinfo{person}{et al.}} \bibinfo{year}{2023}\natexlab{}.
\newblock \showarticletitle{Gender, Smoking History, and Age Prediction from Laryngeal Images}.
\newblock \bibinfo{journal}{\emph{Journal of Imaging}} \bibinfo{volume}{9}, \bibinfo{number}{6} (\bibinfo{year}{2023}), \bibinfo{pages}{109}.
\newblock


\end{thebibliography}
}

\end{document}